\documentclass{appolb}
\usepackage{graphicx}
\usepackage{amsmath}
\usepackage{slashed}
\usepackage{titlesec}
\allowdisplaybreaks

\begin{document}
\title{Exclusive leptoproduction of a light vector meson at the twist-3 in a GTMD framework\thanks{Presented at ``Diffraction and Low-$x$ 2024'', Trabia (Palermo, Italy), September 8-14, 2024.}  }

\author{Renaud Boussarie
\address{CPHT, CNRS, Ecole Polytechnique, Institut Polytechnique de Paris, F-91128 Palaiseau, France}
\\[3mm]
{Michael Fucilla\thanks{Speaker at the conference.}, Samuel Wallon 
\address{Université Paris-Saclay, CNRS/IN2P3, IJCLab, F-91405, Orsay, France}
}
\\[3mm]
Lech Szymanowski
\address{National Centre for Nuclear Research (NCBJ), Pasteura 7, 02-093 Warsaw, Poland}
}

\maketitle
\begin{abstract}
We provide the most general description of the exclusive leptoproduction of a light vector meson, at high center-of-mass energy, within the CGC/shockwave formalism. We keep a twist-3 accuracy in $s$ channel, thus being able to describe all possible helicity amplitudes, including the ones for the production of a transversally polarized meson. In this latter case, we overcome the well-known issue of endpoint singularities by promoting the GPD to a GTMD given by matrix elements of dipole and double dipole operators. The all-twists treatment of the proton (nucleon) target allow to safely twist-expand the general vacuum-to-meson matrix elements. Therefore, unlike previous attempts in the modified perturbative approach, our final results are expressed in terms of the twist-3 collinear distribution amplitudes, introduced in the context of the higher-twist collinear formalism. 
\end{abstract}

\section{Introduction}
The semi-classical effective approach to QCD at small-$x$~\cite{McLerran:1993ni,Balitsky:1995ub} is the established tool for the exploration of semi-hard processes at the Large Hadron Collider (LHC) and at the future Electron-Ion Collider (EIC). Besides the inclusive channels (see for instance~\cite{Caucal:2023fsf,Chirilli:2011km,Beuf:2021srj,Taels:2022tza}), in recent times there has been an increasing attention towards exclusive~\cite{Boussarie:2014lxa,Boussarie:2016ogo,Boussarie:2016bkq,Boussarie:2019ero,Mantysaari:2022kdm,Siddikov:2024bre,Benic:2024pqe} and semi-inclusive~\cite{Marquet:2009ca,Hatta:2022lzj,Iancu:2021rup,Fucilla:2022wcg,Fucilla:2023mkl} diffractive processes. Among them, the electroproduction of a light vector meson is a golden channel and it has been extensively studied at the HERA experiments~\cite{ZEUS:2007iet,H1:2009cml}, where an exhaustive experimental analysis of the spin-density matrix elements of the $\gamma^{(*)} P \rightarrow M (\rho, \phi, \omega) P$ process has been provided. The largest contribution to this spin matrix is the one describing the longitudinal photon to longitudinal meson transition. This latter, from a theoretical point of view, corresponds to the leading twist contribution and the description of all other transitions requires a treatment beyond leading twist. Unfortunately, at the twist-3, the above mentioned process does not factorize in the sense of the standard collinear factorization~\cite{Mankiewicz:1999tt,Anikin:2002wg}. An attempt to solve this problem is the modified perturbative approach~\cite{Goloskokov:2013mba}, in which DAs are promoted to light-cone wavefunctions. Another idea is to move to $k_T$-factorization from the target side. Then, the divergences are regularized by the off-shellness of the gluons in the $t$-channel~\cite{Anikin:2009bf}. We will follow this second approach in the spirit and develop a framework combining the higher-twist formalism of exclusive processes in the $s$-channel with the semi-classical effective description of small-$x$ physics in the $t$-channel. As a result, we will get the result of all helicity amplitudes of the exclusive light vector meson production at small-$x$, including the saturation regime, and in the most general non-forward kinematics. 

\section{Semi-classical small-$x$ formalism with effective operators}
We work within the $k_T$-factorization scheme and introduce a light-cone basis using vectors $n_1$ and $n_2$, which specify the $+/-$ directions and such that 
\begin{equation}
    n_1 \cdot n_2 = 1 \; \hspace{0.5 cm} {\rm and} \; \hspace{0.5 cm} l^{\mu} = l^+ n_1^{\mu} + l^- n_2^{\mu} + l_{\perp} \; ,
\end{equation}
with $l$ a generic four-momenta. The direction of the photon is taken along the $+$-direction, while the one of the incoming proton along the $-$ one. To investigate the light vector meson production at high-energy, we rely on the semi-classical effective approach to QCD at small-$x$~\cite{McLerran:1993ni,Balitsky:1995ub}, in which the gluonic field $\mathcal{A} (k)$ is split into classical background fields $b(k)$ (small $k^+$) and quantum fields $A(k)$ (large $k^+$). Then, in the projectile frame, the target gluon field is effectively treated as a QCD shockwave, i.e.
\begin{equation}
    b^{\mu} (x) = b^{-} (x^+, x_{\perp}) n_2^{\mu} = \delta (x^+) \mathbf{B} \boldsymbol{(x)} n_2^{\mu} \; .
\end{equation}
Multiple re-scattering with the target background field yields a Wilson line located at $x^{-} = 0$ 
\begin{equation*}
    W_{\boldsymbol{z}} = \mathcal{P} {\rm exp} \left( i g \int d z^+ b^{-a} (z) \mathbf{T}^a_r \right) \; , 
\end{equation*}
where $\mathcal{P}$ denotes the path ordering operator in the $+$ direction and $\mathbf{T}^a_r$ is a color matrix in the $r$ representation of the gauge group. In particular, $W_{\boldsymbol{z}} = V_{\boldsymbol{z}} $ ($W_{\boldsymbol{z}} = U_{\boldsymbol{z}}$) in the fundamental (adjoint) representation. \\

Our general aim is to obtain a result expressed in terms of general vacuum-to-meson matrix elements, i.e. without any reference to light-cone expansion at the first stage. To achieve such a description, it is convenient to introduce effective background field operators~\cite{Boussarie:2024bdo,Boussarie:2024pax}, which take into account the infinite eikonal interactions between the projectile and the classical fields at the operator level. We have 
\begin{equation}
    \left[\psi_{\text {eff }}\left(z_0\right)\right]_{z_0^{+}<0} = - \int \mathrm{d}^D z_2 G_0\left(z_{02}\right) V_{\boldsymbol{z}_2}^{\dagger} \gamma^{+} \psi\left(z_2\right) \delta\left(z_2^{+}\right)  \; ,
    \label{Eq:PsiEffecNoMon}
\end{equation}
\begin{equation}
    \left[\bar{\psi}_{\text{eff }}\left(z_0\right)\right]_{z_0^{+}<0} = \int \mathrm{d}^D z_1 \bar{\psi}\left(z_1\right) \gamma^{+} V_{\boldsymbol{z}_1} G_0\left(z_{10}\right) \delta\left(z_1^{+}\right) 
    \label{Eq:PsiBarEffecNoMon}
\end{equation}
and
\begin{equation}
   \left[ A_{\text {eff }}^{\mu a}\left(z_0\right) \right]_{z_0^{+}<0} \hspace{-0.1 cm } = 2 i \hspace{-0.1 cm } \int \hspace{-0.1 cm } \mathrm{d}^D z_3 \delta\left(z_3^{+}\right) F_{-\sigma}^b\left(z_3\right)   G^{\mu \sigma_{\perp}} \hspace{-0.1 cm } \left( z_{30} \right) U_{\boldsymbol{z}_3}^{a b} \; ,
   \label{Eq:AEffecNoMon}
\end{equation}
where $G_0(x)$ is the free quark propagator, $G^{\mu \sigma_{\perp}} \left( x \right)$ is the free gluon propagator in the $n_2$ light-cone gauge and $F_{\mu \nu}^a(x) = \partial_{\mu} A_{\nu}^a (z) - \partial_{\nu} A_{\mu}^a (z) - g f^{abc} A_{\mu}^b (z) A_{\nu}^{c} (z)$ is the QCD field strength tensor. 

\section{Exclusive light vector meson production at the twist-3}
A complete and exhaustive description of the process at the twist-3 requires taking into account both \textit{kinematic} and \textit{genuine} twist effects. The kinematic twist effects are related to the relative transverse motion between the constituents of the leading Fock state wave function. Instead, the genuine twist effects take into account the fact that beyond the leading twist, higher Fock states (non-minimal parton configuration) contribute to the meson wave function. These two effects are physically different, but related to each other by the QCD equations of motion. 

\subsection{2-body contribution: kinematic twist effects}
As anticipated, the kinematic twist contribution requires to calculate the $\gamma^{(*)} P \rightarrow M (\rho, \phi, \omega) P$ amplitude, in which the meson is produced starting from a quark-antiquark pair. Using the effective operator formalism, we obtain  
\begin{gather}
   \mathcal{A}_2 = \int_{0}^{1} {\rm d} x \int{\rm d}^{2} \boldsymbol{r} \Psi_2 \left(x, \boldsymbol{r} \right) \int{\rm d}^{d} \boldsymbol{b} \; {\rm e}^{i (\boldsymbol{q}-\boldsymbol{p}_M) \cdot\boldsymbol{b}} \left\langle P\left(p^{\prime}\right)\left|\mathcal{U}_{\boldsymbol{b}+\overline{x}\boldsymbol{r} \; \boldsymbol{b}-x\boldsymbol{r}} \right|P\left(p\right)\right\rangle \; ,
\label{Eq:StandardDipoleAmp-rb}
\end{gather}
where $q$ $(p_M)$ is the photon (meson) momenta, $x$ is the longitudinal fraction of meson momenta carried by the quark, $\bar{x} \equiv 1-x$, $\mathcal{U}_{\boldsymbol{b}+\overline{x}\boldsymbol{r} \; \boldsymbol{b}-x\boldsymbol{r}}$ is the dipole operator and $\Psi_2 \left(x, \boldsymbol{r} \right)$ is a two-body photon meson wavefunction overlap (PMWO). The explicit expression of $\Psi_2 \left(x, \boldsymbol{r} \right)$ can be found in refs.~\cite{Boussarie:2024bdo,Boussarie:2024pax}, here, we limit to recall that it contains the 2-body vacuum-to-meson matrix elements
\begin{equation}
\phi_{\Gamma^{\lambda}} = \int \frac{ d r^-}{2 \pi} e^{i x \boldsymbol{p}_M \cdot\boldsymbol{r} -i x p_M^+ r^- } \langle M(p_{M}) | \overline{\psi}\left(r\right)\Gamma^{\lambda}\psi\left(0\right) | 0\rangle_{r^+=0} \; ,  
\label{Phi_+_Distrib_Coordinate}
\end{equation}
where $\Gamma^{\lambda}= \{ \gamma^+, \gamma^+ \gamma^5 \}$. The most general expression of $\Psi_2 \left(x, \boldsymbol{r} \right)$ contains an infinite kinematic twist, via the off-light-cone matrix elements in eq.~(\ref{Phi_+_Distrib_Coordinate}). In order to express the final result in terms of collinear twist-3 distribution amplitudes, one has to perform the non-local operator product expansion (OPE)~\cite{Balitsky:1987bk} in powers of the deviation from the light-cone $r^2 = r_{\perp}^2 = 0$. In the case of light vector meson distribution amplitudes, this program was achieved up to twist-4 in ref.~\cite{Ball:1998sk}. Paying attention to some formal subtleties\footnote{In ref.~\cite{Ball:1998sk} the parametrization of the vacuum-to-meson matrix elements are obtain for correlator expressed in the most general gauge invariant form.}, the pioneering formalism of ref.~\cite{Ball:1998sk} can be used to achieve the twist-expansion of (\ref{Phi_+_Distrib_Coordinate}) to finally get
\begin{gather}
     \left[ \Psi_2 \left(x, \boldsymbol{r} \right) \right]_{\rm twist-3} = e_q m_M f_M \delta \left( 1 - \frac{p_M^+}{q^+} \right) \left(\varepsilon_{q\mu}-\frac{\varepsilon_{q}^{+}}{q^{+}}q_{\mu} \right) \left(\varepsilon_{M \alpha}^{*}-\frac{\varepsilon_{M}^{*+}}{p_M^{+}} p_{M \alpha} \right) \nonumber \\ \times \left[ -i r_{\perp}^{\alpha} (h(x) - \tilde{h}(x)) \left( 2 x \bar{x} q^{\mu} + (x-\bar{x}) \frac{- i \partial}{ \partial r_{\perp \mu}} \right) \right. \nonumber \\ \left. + \epsilon^{\mu \nu + -} \epsilon^{+ \alpha - \delta} r_{\perp \delta} \left( \frac{g^{(a)}_{\perp} (x) - \tilde{g}_{\perp}^{(a)} (x)}{4} \right) \frac{\partial}{\partial r_{\perp}^{\nu}} \right] K_0 \left( \sqrt{x \bar{x} Q^2 \boldsymbol{r}^{2}} \right) ,
     \label{Eq:PMOW_2Body_Twist-3}
\end{gather}
where $m_M$ is the meson mass, $e_q$ the quark electric charge, $\varepsilon_{q}$ ($\varepsilon_{M}^{*}$) is the polarization vector of the photon (meson) and $f_M$ is the standard vector decay constant~\cite{Ball:1998sk}. Moreover, $h(x)$ and $g^{(a)}_{\perp} (x)$ are kinematic twist-3 DAs while 
\begin{gather}
  \widetilde{h}\left(x\right)=\frac{f_{3M}^{V}}{f_{M}}\int_{0}^{x}{\rm d}x_{q}\int_{0}^{1-x}\!{\rm d}x_{\overline{q}}\,\frac{V\left(x_{q},x_{\overline{q}}\right)}{\left(1-x_{q}-x_{\overline{q}}\right)^{2}} \; , \nonumber \\ \widetilde{g}_{\perp}^{\left(a\right)}\left(x\right)=4\frac{f_{3M}^{A}}{f_{M}}\int_{0}^{x}{\rm d}x_{q}\int_{0}^{1-x}\!{\rm d}x_{\overline{q}}\,\frac{A\left(x_{q},x_{\overline{q}}\right)}{(1-x_{q}-x_{\overline{q}}+i\epsilon)^{2}} \; , 
\end{gather}
where $V(x_1, x_2)$ ($A(x_1, x_2)$) is the genuine vector (axial) twist-3 DA. $f_M^V$ and $f_M^A$ are the vector and axial normalization constants~\cite{Ball:1998sk}.

\subsection{3-body contribution: genuine twist effects}
The genuine twist contribution requires to calculate the $\gamma^{(*)} P \rightarrow M (\rho, \phi, \omega) P$ amplitude, in which the meson is produced starting from a quark-antiquark-gluon system. In this case the factorization formula is
\begin{gather}
    \mathcal{A}_3 = \left( \prod_{i=1}^3 \int \hspace{-0.1 cm} d x_i \hspace{-0.1 cm} \int \hspace{-0.1 cm} d^2 \boldsymbol{z}_i e^{i x_i \boldsymbol{q} \boldsymbol{z}_i} \theta (x_i) \right) \hspace{-0.1 cm} \delta (1 - \sum_i x_i )
    \Psi_3 \left( \{ x_i \} , \{ \boldsymbol{z}_i \} \right) \nonumber \\ \times \left\langle P\left(p^{\prime}\right)\left| \mathcal{U}_{\boldsymbol{z}_1 \boldsymbol{z}_3} \mathcal{U}_{\boldsymbol{z}_3 \boldsymbol{z}_2} - \mathcal{U}_{\boldsymbol{z}_1 \boldsymbol{z}_3} -\mathcal{U}_{\boldsymbol{z}_3 \boldsymbol{z}_2} + \frac{ \mathcal{U}_{\boldsymbol{z}_1 \boldsymbol{z}_2}}{N_c^2}  \right|P\left(p\right)\right\rangle ,
   \label{Eq:GenealStructure3bodyWaveFunOver}
\end{gather}
where $\mathcal{U}_{\boldsymbol{z}_1 \boldsymbol{z}_3} \mathcal{U}_{\boldsymbol{z}_3 \boldsymbol{z}_2}$ is the double dipole operator,
$ \Psi_3 \left( \{ x_i \} , \{ \boldsymbol{z}_i \} \right)$ is a 3-body PMWO and $\{ a_i \} = \{ a_1, a_2, a_3 \}$. Again the expression of $ \Psi_3 \left( \{ x_i \} , \{ \boldsymbol{z}_i \} \right) $ can be found in refs.~\cite{Boussarie:2024bdo,Boussarie:2024pax} and it depends on the 3-body vacuum-to-meson matrix elements, i.e.
\begin{gather}
    \chi_{\Gamma^{\lambda} , \sigma } 
   = \hspace{-0.1 cm} \int_{-\infty}^{\infty} \hspace{-0.1 cm} \frac{ {\rm d} z_{1}^{-}}{2 \pi} \frac{ {\rm d} z_{2}^{-}}{2 \pi} \frac{ {\rm d} z_{3}^{-}}{2 \pi} {\rm e}^{-ix_{1}q^{+}z_{1}^{-}-ix_{2}q^{+}z_{2}^{-}-ix_{3}q^{+}z_{3}^{-}} \nonumber \\ \times \left\langle M\left(p_{M}\right)\left|\overline{\psi}\left(z_{1}\right) g \Gamma^{\lambda} F_{-\sigma}\left(z_{3}\right)\psi\left(z_{2}\right)\right|0\right\rangle _{z_{1,2,3}^{+}=0} .
   \label{Eq:Chi+_Vector_function}
\end{gather}
Performing the twist-3 expansion of $\Psi_3 \left( \{ x_i \} , \{ \boldsymbol{z}_i \} \right)$ is rather straightforward since, within this accuracy, the correlators (\ref{Eq:Chi+_Vector_function}) coincide with those in~\cite{Ball:1998sk}. We do not report the explicit expression for compactness. It can be found in refs.~\cite{Boussarie:2024bdo,Boussarie:2024pax} and is a function of the genuine twist-3 DAs $A(x_1,x_2)$ and $V(x_1, x_2)$, introduced previously.   

\section{Summary and conclusion}
We introduced a general framework to deal with beyond leading power corrections at small-$x$ and applied it to the transversely polarized light vector meson production, $\gamma^{(*)} P \rightarrow M (\rho, \phi, \omega) P$, which starts at the next-to-leading power and for which a purely collinear treatment leads to end-point singularities. Existing and future measurements of the whole spin density matrix requires to go beyond leading twist, since some of these matrix elements simply vanish at leading twist. Considering higher-twist corrections in the high-energy formalism has several advantages. First, at hard scale values close to the saturation scale, exclusive or semi-inclusive processes can be strongly sensitive to higher-twist effects. Taking their effect into account is important in view of precision studies. As a by-product, a (generalized) $k_T$-dependent factorization, reliable at high-energy, is a valid option to investigate physical processes for which collinear factorization breaks down, as the one considered here.

\section{Acknowledgments}
This  project  has  received  funding  from  the  European  Union’s  Horizon  2020  research  and  innovation program under grant agreement STRONG–2020 (WP 13 "NA-Small-x"). The work by M.~F. is supported by Agence Nationale de la Recherche under the contract ANR-17-CE31-0019. M. F. acknowledges support from the Italian Foundation “Angelo Della Riccia”. The  work of  L.~S. is  supported  by  the  grant  2019/33/B/ST2/02588  of  the  National  Science Center in  Poland. L.~S. thanks the P2IO Laboratory of Excellence (Programme Investissements d'Avenir ANR-10-LABEX-0038) and the P2I - Graduate School of Physics of Paris-Saclay University for support. This work was also partly supported by the French CNRS via the GDR QCD.

\end{document}